\documentstyle[aps,twocolumn,epsfig]{revtex}

\begin{document}
\draft

\twocolumn[\hsize\textwidth\columnwidth\hsize\csname@twocolumnfalse\endcsname
\title{Pairwise entanglement in symmetric multi-qubit systems}
\author{Xiaoguang Wang and Klaus M\o lmer}
\address{Institute of Physics and Astronomy, University of Aarhus, DK-8000, Aarhus C, 
Denmark.}
\date{\today}
\maketitle

\begin{abstract}
The concurrence, a quantitative measure of the entanglement
between a pair of particles, is determined for
the case where the pair is extracted from a symmetric 
state of $N$ two-level systems.  Examples are given
for both pure and mixed states of the $N$-particle
system, and for a pair extracted from two ensembles with
correlated  collective spins.
\end{abstract}

] \narrowtext

\section{Introduction}

Various proposals exist for the preparation of multi-particle entangled
states, and a number of these states have been pointed out to be particularly
easy to prepare and to have special and useful properties. Since
entanglement is defined as a property of the whole ensemble of particles, it
is not immediately clear whether a sub-ensemble of particles, drawn at
random from the original ensemble will also be in an entangled state, or
whether the trace over some particles will destroy the quantum correlations
in the system. In this paper we consider the simple questions whether a
random pair of particles, extracted from a symmetric state of $N$ two-level
systems will be in an entangled state or not, where by symmetric, we assume
symmetry under any permutation of the particles. Entangled states constitute
a valuable resource in quantum information processing\cite{Bennett}, and the
transfer of entanglement between few qubits and the quasi-continuous
variables by which we describe many-particle systems, may become an
important ingredient in, e.g., quantum data-storage and inter-species
teleportation.

The paper is organized as follows. In Sec. II, we present the concurrence,
introduced by Wootters \cite{wootters1,wootters2}, who demonstrated its
one-to-one correspondence with the entanglement of formation of a pair of
qubits. In Sec. III, we show how  the density matrix of a pair of qubits
can be expressed in terms of expectation values of collective spin operators
on the multi-qubit state. In Sec. IV, we analyze three examples 
of pure states of the $N$ particles: spin coherent
states, Dicke states, and spin squeezed states. In Sec. V, we consider an
example of a mixed state with thermal entanglement \cite{thermal,Wang1,Wang2,Ising,Wang3}, and we
show examples where the pairwise entanglement depends on the temperature of
the system.
Finally in Sec. VI, we assume two separate ensembles in an 
Einstein-Podolsky-Rosen state of correlated angular momentum components, and we show
that a single pair with an atom from each ensemble will be in an entangled
state.

\section{Two-particle density matrices and entanglement}

It is easy to check if a pure state of two quantum systems is an entangled
state or not by simply observing the eigenvalues $r_i$ of the reduced
density matrix of either system. It is also possible to quantity the amount
or degree of entanglement of the state \cite{Bennett96}, $E=-\sum_i r_i
\log_2r_i$, which presents the asymptotic ratio between $n$ and $m$, where $%
n $ is the number of pairs in the desired state, synthesized from $m$ pairs
of maximally entangled states.

For a mixed state with density matrix $\rho _{12}$, a similar measure can be
defined as the minium value of the weighted average of $E$ over wave
functions by which the two-particle density matrix can be written as a
weighted sum. It is necessary to search for the minimum, since $\rho _{12}$
can be written in many ways as a weighted sum of pure state projections. In
the general case, this is a highly non-trivial task, as is the determination
whether the state is entangled at all. For two qubits, however, entanglement
is equivalent with the non-positivity of the partially transposed density
matrix \cite{peres}, and the entropy of formation can, magically, be
obtained as a simple analytical expression \cite{wootters1,wootters2} 
\begin{equation}
E=h\left( \frac{1+\sqrt{1-{\cal C}^2}}2\right)   \label{entanglement}
\end{equation}
where $h(x)=-x\log _2x-(1-x)\log _2(1-x)$, and where the {\it concurrence}, $%
{\cal C}$, is defined as 
\begin{equation}
{\cal C}=\max \left\{0, \lambda _1-\lambda _2-\lambda _3-\lambda _4\right\} ,
\label{Cdef}
\end{equation}
where the quantities $\lambda _i$ are the square roots of the eigenvalues 
in descending order of
the matrix product 
\begin{equation}
\varrho _{12}=\rho _{12}(\sigma _{1y}\otimes \sigma _{2y})\rho
_{12}^{*}(\sigma _{1y}\otimes \sigma _{2y}).  \label{varrho}
\end{equation}
In (\ref{varrho}) $\rho _{12}^{*}$ denotes the  complex
conjugate of $\rho _{12}$, and $\sigma _{iy}$ are Pauli matrices for the
two-level systems. The eigenvalues of $\varrho _{12}$ are real and
non-negative even though $\varrho _{12}$ is not necessarily Hermitian, and
the values of the concurrence range from zero for an unentangled state to
unity for a maximally entangled state.

\section{Density matrix for a pair of qubits from a multi-qubit state}

A two-qubit reduced density matrix which is symmetric under exchange of the two
systems can be written as
\begin{equation}
\rho _{12}=\left( 
\begin{array}{llll}
v_{+} & x_{+}^{*} & x_{+}^{*} & u^{*} \\ 
x_{+} & w & y^{*} & x_{-}^{*} \\ 
x_{+} & y & w & x_{-}^{*} \\ 
u & x_{-} & x_{-} & v_{-}
\end{array}
\right)   \label{eq:rhogood}
\end{equation}
where the matrix elements in the basis $\{|00\rangle ,|01\rangle ,|10\rangle
,|11\rangle \}$ can be represented by expectation values of Pauli spin
matrices of the two systems 
\begin{eqnarray}
v_{\pm } &=&\frac 14\left( 1\pm 2\langle \sigma _{1z}\rangle +\langle \sigma
_{1z}\sigma _{2z}\rangle \right) , \nonumber \\
x_{\pm } &=&\frac 12(\langle \sigma _{1+}\rangle \pm \langle \sigma
_{1+}\sigma _{2z}\rangle ), \nonumber \\
w &=&\frac 14\left( 1-\langle \sigma _{1z}\sigma _{2z}\rangle \right) , \nonumber \\
y&=&\langle\sigma_{1+}\sigma_{2-}\rangle, \nonumber\\
u &=&\frac 14(\langle \sigma _{1x}\sigma _{2x}\rangle -\langle \sigma
_{1y}\sigma _{2y}\rangle +i2\langle \sigma _{1x}\sigma _{2y}\rangle ).
\label{eq:para1}
\end{eqnarray}

We now consider the entanglement of two qubits extracted from a symmetric
multi-qubit states. If only symmetric qubit states are considered, we can
describe the state of the $N$-qubit system in terms of the orthonormal basis $%
|S,M\rangle (M=-S,-S+1,...,S)$ with $S=N/2$. The states $|S,M\rangle $ are
the usual symmetric Dicke state\cite{Dicke}, i.e., eigenstates of the
collective spin operators $\vec{S}^2$ and $S_z$, defined as 
\begin{equation}
S_\alpha =\frac 12\sum_{i=1}^N\sigma _{i\alpha },\ \ \ \alpha =x,y,z.
\end{equation}

For later use it is convenient to define the number operator ${\cal N}%
=S_z+N/2$ and number states as 
\begin{eqnarray}
|n\rangle _N &\equiv &|N/2,-N/2+n\rangle _N,  \nonumber \\
{\cal N}|n\rangle _N &=&n|n\rangle _N.
\end{eqnarray}
The eigenvalue $n$ of the number operator ${\cal N}$ is the number of qubits
in the state $|0\rangle $. For example, the states $|0\rangle _N$ and $%
|1\rangle _N$ are explicitely written as 
\begin{eqnarray}
|0\rangle _N &=&|111...1\rangle , \\
|1\rangle _N &=&\frac 1{\sqrt{N}}(|011...1\rangle +|101..1\rangle   \nonumber
\\
&&+...+|111...0\rangle ). \label{eq:w}
\end{eqnarray}
$|1\rangle _N$ is also called an $N$--qubit ${\rm W}$ state\cite{Dur00,Wang1}%
.

Due to the symmetry of the state under exchange of particles we have 
\begin{eqnarray}
\langle \sigma _{1\alpha }\rangle  &=&\frac{2\langle S_\alpha \rangle }N, \label{eq:re1}\nonumber\\
\langle \sigma _{1+}\rangle  &=&\frac{\langle S_{+}\rangle }N, \nonumber \\
\langle \sigma _{1\alpha }\sigma _{2\alpha }\rangle  &=&\frac{4\langle
S_\alpha ^2\rangle -N}{N(N-1)}, \nonumber\\
\langle \sigma _{1x}\sigma _{2y}\rangle  &=&\frac{2\langle
[S_x,S_y]_{+}\rangle }{N(N-1)}, \nonumber\\
\langle \sigma _{1+}\sigma _{2z}\rangle  &=&\frac{\langle
[S_{+},S_z]_{+}\rangle }{N(N-1)},  \label{eq:relation}
\end{eqnarray}
where $[A,B]_{+}=AB+BA$ is the anticommutator for operators $A$ and $B.$

From Eqs.(\ref{eq:para1}) and (\ref{eq:relation}) we may thus express the
density matrix elements of $\rho _{12}$ in terms of the expectation values
of the collective operators, 
\begin{eqnarray}
v_{\pm } &=&\frac{N^2-2N+4\langle S_z^2\rangle \pm 4\langle S_z\rangle (N-1)%
}{4N(N-1)},  \label{eq:aaa} \nonumber \\
x_{\pm } &=&\frac{(N-1)\langle S_{+}\rangle \pm \langle
[S_{+},S_z]_{+}\rangle }{2N(N-1)}, \nonumber \\
w &=&\frac{N^2-4\langle S_z^2\rangle }{4N(N-1)}, \nonumber \\
y &=& \frac{2\langle S_x^2+S_y^2\rangle -N}{2N(N-1)}, \nonumber\\
u &=&\frac{\langle S_x^2-S_y^2\rangle +i\langle [S_x,S_y]_{+}\rangle }{N(N-1)%
}=\frac{\langle S_{+}^2\rangle }{N(N-1)}.  \label{eq:bbb}
\end{eqnarray}

\section{Pure multiqubit states}

In this section we study three examples, where the $N$ two-level systems are 
described by a pure state which is invariant under permutation of
the particles.
\subsection{Spin coherent states}

The spin coherent state \cite{SCS} is obtained by a rotation of the spin state $%
|S,M=S\rangle $, which in turn is the product state of all $N$ particles
in the $|0\rangle $ state. Hence it is a separable state. It is still
interesting to go through the above procedure and to insert the explicit
expression of the spin coherent state \cite{SCS}, 
\begin{equation}
|\eta \rangle =(1+|\eta |^2)^{-N/2}\sum_{n=0}^N\left( 
\begin{array}{c}
N \\ 
n
\end{array}
\right) ^{1/2}\eta ^n|n\rangle _N,  \label{eq:scs}
\end{equation}
where $\eta $ is chosen real in the following. 
By a straightforward calculation from Eqs.(\ref{eq:bbb}) and (\ref{eq:scs}), we find 
\begin{equation}
\rho _{12}=\frac 1{(1+\eta ^2)^2}\left( 
\begin{array}{llll}
\eta ^4 & \eta ^3 & \eta ^3 & \eta ^2 \\ 
\eta ^3 & \eta ^2 & \eta ^2 & \eta  \\ 
\eta ^3 & \eta ^2 & \eta ^2 & \eta  \\ 
\eta ^2 & \eta  & \eta  & 1
\end{array}
\right) 
\end{equation}
which is in agreement with our observation that the two-particle state is
really a product state of two rotated spin-$\frac 12$ particles in the
states $(\eta |0\rangle +|1\rangle )/\sqrt{{1+\eta ^2}}$. The matrix product 
$\varrho _{12}$ is found to be a 4$\times 4$ matrix of zero's, revealing the
role of the $\sigma _y$ Pauli matrices in (\ref{varrho}): $\rho _{12}$ is the
projection operator on spin states with a definite direction in the $xz$%
-plane, the application of $\sigma _y$ is equivalent to a $180^{\circ }$
rotation in the $xz$-plane, and $\varrho _{12}$ is therefore the vanishing
product of projection operators on two orthogonal subspaces. Naturally, the
concurrence vanishes in this case, ${\cal C}=0$; there is no
pairwise entanglement in the spin coherent state. 

\subsection{Dicke State $|N/2,M\rangle $}

The Dicke states, defined as effective number states above, are states with
a definite number of particles occupying the internal states $|0\rangle $
and $|1\rangle $. Such states may in principle be prepared in an atomic
physics experiment by Quantum Non-Demolition detection of the atomic
populations by phase contrast imaging of the atomic sample \cite
{klaus,kuzmich}. By rotation of all spins, a separable spin coherent state
is first prepared with a binomial distribution on the various Dicke states, cf.,
Eq.(\ref{eq:scs}), and experiments have already demonstrated a factor 3
reduction in the variance of the populations after such a detection 
\cite{kuzmich2}.

From Eq.(\ref{eq:bbb}), it is easy to see that the
reduced density matrix $\rho _{12}$ is given by 
\begin{equation}
\rho _{12}=\left( 
\begin{array}{llll}
v_+ & 0 & 0 & 0 \\ 
0 & w & w & 0 \\ 
0 & w & w & 0 \\ 
0 & 0 & 0 & v_-
\end{array}
\right)   \label{eq:rho}
\end{equation}
with matrix elements 
\begin{eqnarray}
v_{\pm } &=&\frac{(N\pm 2M)(N-2\pm 2M)}{4N(N-1)},  \nonumber \\
w &=&\frac{N^2-4M^2}{4N(N-1)}.  \label{eq:elements}
\end{eqnarray}
The concurrence of a simple density matrix of the form (\ref
{eq:rhogood}) with $x_{\pm}, u=0,$ and $y=y^*$ is given by\cite{Ring} 
\begin{eqnarray}
{\cal C} &=&2\max \{0,y-\sqrt{v_{+}v_{-}}\}, \label{eq:c1}
\end{eqnarray}
where we have used that $2y+v_{+}+v_{-}=1$ to $\rho_{12}$. 
Now substituting Eq.(\ref{eq:elements}) into (\ref{eq:c1}), we explicitly obtain 
\begin{eqnarray}
{\cal C} &=&\frac 1{2N(N-1)}\{N^2-4M^2  \nonumber \\
&&-\sqrt{(N^2-4M^2)[(N-2)^2-4M^2]}\}.
\end{eqnarray}
The values of ${\cal C}$ for different $N$ and $M$ are illustrated in Fig.
1. For any Dicke state except the ones with maximum $|M|$, if one extracts
two particles, they will be in an entangled state. We also observe that
the concurrence is nearly a constant in the neighborhood of $M=0$.

\begin{figure}[tbh]
\begin{center}
\epsfxsize=9cm
\epsffile{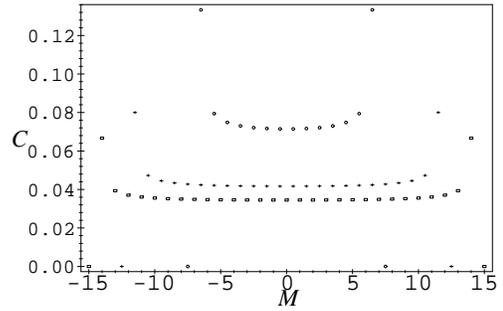}
\end{center}
\caption{The concurrence in the Dicke state for different number $N$. 
$N=15$(open circles), $N=25$ (crosses), and $N=30$ (open square).}
\end{figure}

The variation of $M$ around $M=0$ is small for an initial binomial
distribution with this mean value, and the concurrence will be very close to
the exact result, ${\cal C}=1/(N-1)$ for $M=0$, irrespective of the outcome
of a QND measurement of $M$.

The Dicke states $|N/2,M=\pm (N/2-1)\rangle$ have a concurrence of ${\cal C}
=2/N$. These states are identical with the {\rm W} state (see Eq.(\ref{eq:w})), 
which are known to be the symmetric states with the highest
possible concurrence \cite{koashi}.

\subsection{Kitagawa-Ueda state}

In 1993, Kitagawa and Ueda proposed a nonlinear Hamiltonian $\chi S_x^2$ in
order to generate spin squeezed states\cite{Kitagawa}. This effective
Hamiltonian may be realized in ion traps \cite{ions}, where it was already
implemented in order to produce multi-particle entangled states (of four
particles) \cite{sackett}, and it may be implemented in two-component
Bose-Einstein condensates as a direct consequence of the collisional
interactions between the particles \cite{ander_nature}, see also \cite{uffe}.

When the Hamiltonian $H=\chi S_x^2$ is applied to the many-particle system,
which has been prepared in the product state $|0\rangle _N=|111,...,1\rangle 
$, the wave function at time $t$ is obtained as 
\begin{equation}
|\Psi (t)\rangle =e^{-i\chi tS_x^2}|0\rangle _N.
\end{equation}
Using the results obtained in \cite{Kitagawa} the following expectation
values are obtained ($\mu =2\chi t$) 

\begin{eqnarray}
\langle S_x\rangle  &=&\langle S_y\rangle =0,  \label{eq:aaaa} \nonumber \\
\langle S_z\rangle  &=&-\frac N2\cos ^{N-1}\left( \frac \mu 2\right)  \nonumber  \\
\langle S_x^2\rangle  &=&N/4 \nonumber  \\
\langle S_y^2\rangle  &=&\frac 18\left( N^2+N-N(N-1)\cos ^{N-2}\mu \right) 
\nonumber \\
\langle S_z^2\rangle  &=&\frac 18\left( N^2+N+N(N-1)\cos ^{N-2}\mu \right) 
\nonumber \\
\langle [S_{+},S_z]_{+}\rangle  &=&0 \nonumber \\
\langle [S_x,S_y]_{+}\rangle  &=&\frac 12N(N-1)\cos ^{N-2}\frac \mu 2\sin 
\frac \mu 2.  \label{eq:bbbb}
\end{eqnarray}

We are now able to determine the two-particle density matrix, which is on
the form 
\begin{equation}
\rho _{12}=\left( 
\begin{array}{llll}
v_{+} & 0 & 0 & u^{*} \\ 
0 & w & w & 0 \\ 
0 & w & w & 0 \\ 
u & 0 & 0 & v_{-}
\end{array}
\right)   \label{eq:rhorho}
\end{equation}
with matrix elements given by Eq.(\ref{eq:bbb}). The
combination of Eqs.(\ref{eq:bbb}) and (\ref
{eq:bbbb}) gives explicitly the matrix elements.

From Eqs.(\ref{Cdef}) and (\ref{varrho}), the concurrence
for the matrix (\ref{eq:rhorho}) is obtained as
\begin{equation}
{\cal C}=\left\{ 
\begin{array}{c}
2\max (0,|u|-w),\text{ if }2w<\sqrt{v_{+}v_{-}}+|u|; \\ 
2\max (0,w-\sqrt{v_{+}v_{-}}),\text{ if }2w\geq \sqrt{v_{+}v_{-}}+|u|.
\end{array}
\right.   \label{eq:c2}
\end{equation}

The concurrence of the spin squeezed states is given by analytical
expressions in the argument $\mu=2\chi t$, which are too lengthy to present
her. In Fig. 2 we present the results numerically: If two atoms are
extracted at random from spin squeezed samples they will be in a mutually
entangled state. We observe that the concurrence is symmetric with
respect to $\mu=\pi$. At this special point of $\mu=\pi$ the $N$--particle GHZ
state is produced\cite{ions}, and it has no pairwise entanglement. 

\begin{figure}[tbh]
\begin{center}
\epsfxsize=9cm
\epsffile{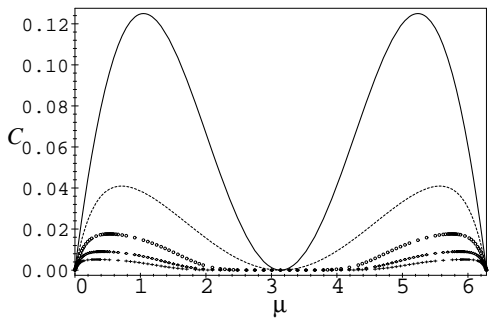}
\end{center}
\caption{The concurrence as a function of  $\mu$ for different number $N$. 
$N=3$(solid line), $N=4$ (dashed line), $N=5$ (open circles), 
$N=6$ (open diamonds), and $N=7$ (crosses).}
\end{figure}

\section{Mixed multiqubit states and thermal entanglement}

An interesting and novel type of thermal entanglement was introduced and
analyzed within the Heisenberg $XXX$\cite{thermal}, $XX$\cite{Wang1}, and $%
XXZ$\cite{Wang2} models as well as within the Ising model in a magnetic field%
\cite{Ising}. The state of the system at thermal equilibrium is represented
by the density operator $\rho (T)=\exp \left( -H/kT\right) /Z,$ where $Z=$tr$%
\left[ \exp \left( -H/kT\right) \right] $ is the partition function, $H$ the
system Hamiltonian, $k$ is Boltzmann's constant which we henceforth take
equal to unity, and $T$ the temperature. As $\rho (T)$ represents a thermal
state, the entanglement in the state is called {\em thermal entanglement}%
\cite{thermal}. Unlike in standard statistical physics where all properties
are obtained from the partition function, determined by the eigenvalues of
the system, entanglement properties require in addition knowledge of the
eigenstates. The analytical results in the previous studies on thermal
entanglement are only available for two\cite{thermal,Wang1,Wang2,Ising} 
and three qubits\cite{Wang3}. Here we consider
pairwise entanglement in the multiqubit systems.

\subsection{Isotropic Heisenberg model}

We consider the $N$--qubit isotropic Heisenberg Hamiltonian 
\begin{equation}
H_I=\frac J4\sum_{i\neq j}^N\left( \sigma _i^x\sigma _j^x+\sigma _i^y\sigma
_j^y+\sigma _i^z\sigma _j^z\right)   \label{eq:h}
\end{equation}
The positive (negative) $J$ corresponds to the antiferromagnetic
(ferromagnetic) case. In this model all particles interact with
each other.

By using the collective spin operators, the Hamiltonian is rewritten as 
\begin{equation}
H=J\left( S_x^2+S_y^2+S_z^2\right) =J\vec{S}^2  \label{eq:hh}
\end{equation}
up to a trivial constant.

Unlike pure states, the symmetric multi-particle density matrix does not
only populate the fully symmetric Dicke states, and we have to determine the
number of collective spin-$S$ states for each $S$. Write $S$ as $(N/2-k)$,
we know that for $k=0$, a single irreducible representation exists: the $N+1$
fully symmetric Dicke states with $S=N/2$. The number of irreducible
representation with $S=N/2-1$ is obtained by noting that their maximum $M$
value is also $N/2-1$, and a total of $\left( 
\begin{array}{c}
N \\ 
1
\end{array}
\right) = N$ states exist with precisely one particle in the $|1\rangle$
state. One of these belong to the $S=N/2$ irreducible representation, and
the remaining $N-1$ states must have $S=N/2-1$. This argument can now be
repeated to obtain the number of states with $S=N/2-2$ and $M=N/2-2$, i.e.,
the number of $S=N/2-2$ irreducible representation, etc., until all $2^N$
states of the system have been accounted for.

The isotropic Hamiltonian only depends on $\vec{S}^2$, and knowing the
multiplicity of each value of this quantity we write the partition function 
\begin{equation}
Z=\sum_{k=0}^{N/2}N_k[2(N/2-k)+1]e^{-\beta J(N/2-k)(N/2-k+1)},  \label{Z}
\end{equation}
where $N_k=\left( 
\begin{array}{c}
N \\ 
k
\end{array}
\right) - \left( 
\begin{array}{c}
N \\ 
k-1
\end{array}
\right)$ follows from the above argument. We assume $\left( 
\begin{array}{c}
N \\ 
-1
\end{array}
\right) =0.$

The reduced density matrix for two qubits is
\begin{equation}
\rho _{12}=\left( 
\begin{array}{llll}
v & 0 & 0 & 0 \\ 
0 & w & y & 0 \\ 
0 & y & w & 0 \\ 
0 & 0 & 0 & v
\end{array}
\right) 
\end{equation}
with matrix elements given by Eq.(\ref{eq:bbb}). The 
matrix element $v=v_{\pm}$ since $\langle S_z\rangle=0$.

Due to the symmetry property $\langle S_x^2\rangle=\langle S_y^2\rangle
=\langle S_z^2\rangle$, we only need to determine 
\begin{eqnarray}
\langle S_z^2\rangle  &=&\sum_{k=0}^{N/2}N_k\sum_{m=0}^{N-2k}(m-N/2+k)^2 \\
&&e^{-\beta J(N/2-k)(N/2-k+1)}/Z.  \nonumber
\end{eqnarray}
And from Eq.(\ref{eq:c1}), the concurrence is obtained as 
\begin{eqnarray}
{\cal C} 
&=&\frac 1{2N(N-1)}\max \{0,  \nonumber \\
&&2|2\langle S_x^2+S_y^2\rangle -N|-N^2+2N-4\langle S_z^2\rangle \}\nonumber\\
&=&\frac 1{2N(N-1)}\max \{0,  \nonumber \\
&&2|4\langle S_z^2\rangle -N|-N^2+2N-4\langle S_z^2\rangle \}.
\label{Ctemp}
\end{eqnarray}

To identify the sign of $A\equiv 2|4\langle S_z^2\rangle -N|-N^2+2N-4\langle
S_z^2\rangle $ in (\ref{Ctemp}), we consider the case where $4\langle
S_z^2\rangle \geq N,$ for which $A=4\langle S_z^2\rangle -N^2$.
Since $%
\langle S_z^2\rangle \leq \frac 13\frac N2\left( \frac N2+1\right) <\frac{N^2%
}4$, we always have $4\langle S_z^2\rangle -N^2<0$, and there is no pairwise
entanglement. 
In the opposite case where $4\langle S_z^2\rangle <N,$ we have $%
A=4N-12\langle S_z^2\rangle -N^2$,
since  $\langle S_z^2\rangle \geq 0$, we have $A\le 0$ if $N\geq 4$.
For case of $N=3$, we have shown that the pairwise thermal entanglement
is absent from both the antiferromagnetic and ferromagnetic isotropic
model\cite{Wang3}.

So we conclude that there is no thermal entanglement for $N\geq 3$ in the
isotropic Heisenberg model. The case of $N=2$ is discussed in detail
in Ref.\cite{thermal} and it is shown that there is no thermal entanglement for
the ferromagnetic case. In order to observe the pairwise entanglement in the 
multiqubit system, now we consider the anisotropic Heisenberg model.

\subsection{Anisotropic Heisenberg model}

The anisotropic Heisenberg Hamiltonian  is given by
\begin{equation}
H_a=J\left( S_x^2+S_y^2+\Delta S_z^2\right) =J\vec{S}^2+J(\Delta -1)S_z^2,
\end{equation}
where $\Delta $ is the anisotropy parameter. Obviously the Hamiltonian $H_a$
reduces to $H_I$ when $\Delta =1,$ and $H_a$ yields the $XX$ model when $\Delta
=0.$ 

The concurrence is still given by (\ref{Ctemp}), but the partition function
and the relevant expectation values now become

\begin{eqnarray}
Z &=&\sum_{k=0}^{N/2}N_k\sum_{m=0}^{N-2k}e^{-\beta J(\Delta -1)(m-N/2+k)^2}
\\
&&\times e^{-\beta J(N/2-k)(N/2-k+1)},  \nonumber \\
\langle S_z^2\rangle  &=&\sum_{k=0}^{N/2}N_k\sum_{m=0}^{N-2k}(m-N/2+k)^2 \nonumber \\
&&\times e^{-\beta J(\Delta -1)(m-N/2+k)^2} \nonumber \\
&&\times e^{-\beta J(N/2-k)(N/2-k+1)}/Z,  \\
\langle S_x^2+S_y^2\rangle 
&=&\sum_{k=0}^{N/2}N_k\sum_{m=0}^{N-2k}[(N/2-k)(N/2-k+1) \nonumber \\
&&-(m-N/2+k)^2] \nonumber \\
&&\times e^{-\beta J(\Delta -1)(m-N/2+k)^2}\text{ } \nonumber \\
&&\times e^{-\beta J(N/2-k)(N/2-k+1)}/Z.
\end{eqnarray}

This model leads to pairwise entanglement, as shown by the numerical results
presented in Figure 3 as functions of the reciprocal temperature, $x=\beta J$%
. For $N=2$ we observe that the concurrence is symmetric with respect to 
$x=0$, which is consistent with the result in Ref.\cite{Wang1}. In other words,
the thermal entanglement appears for both the antiferromagnetic and ferromagnetic
cases. However for $N\ge 3$, the thermal entanglement only exists for the ferromagnetic case.
We observe a critical value of $x$, after which the entanglement vanishes. And the critical value
increases as $N$ increases.

\begin{figure}[tbh]
\begin{center}
\epsfxsize=9cm
\epsffile{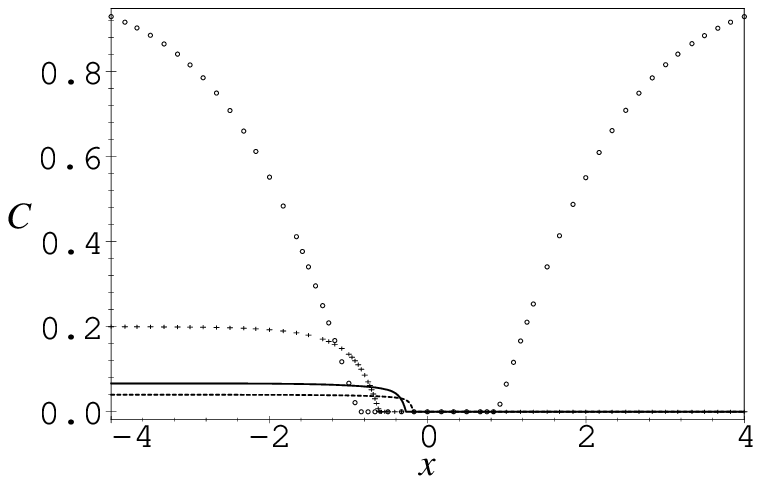}
\end{center}
\caption{The concurrence as a function of  $x=\beta J$ for different number $N$ in the $XX$ model 
($\Delta=0$): $N=2$(open circle), $N=5$ (crosses), $N=15$ (solid line), 
and $N=25$ (dashed line).}
\end{figure}

Within the above framework we may also consider more general models such as
\begin{equation}
H_g=J\vec{S}^2+f(S_z),
\end{equation}
where $f(S_z)$ is an arbitrary analytical function of $S_z$. As the operator $f(S_z)$ commutes with
$\vec{S}^2$, similar analytical results for the concurrence can be obtained and the thermal entanglement
can be readily generated for special choices of $f(S_z)$.

\section{EPR-correlated ensembles}

Finally we consider two EPR-correlated ensembles. This state is not invariant under any permutation of particles,
but only under those permutations that exchange particles within each ensemble, and it is furthermore characterized
by the correlations between the samples 1 and 2:
\begin{eqnarray}
(J_{1x}-J_{2x})|\Psi \rangle &=&0, \label{eq:jj1}\\ 
(J_{1y}+J_{2y})|\Psi \rangle &=&0. \label{eq:jj2}
\end{eqnarray}
A state that obeys Eqs.(\ref{eq:jj1}) and (\ref{eq:jj2}) can 
in principle be obtained by successive QND detection of the observables
$J_{1x}-J_{2x}$ and $J_{1y}+J_{2y}$\cite{Duan,Polzik}.
Equivalently the above equations can be written as
\begin{eqnarray}
(J_{1+}-J_{2-})|\Psi \rangle &=&0, \\
(J_{1-}-J_{2+})|\Psi \rangle &=&0.
\end{eqnarray}
It is easy to check that a solution of the above equation is
the EPR-correlated state
\begin{equation}
|\Psi \rangle =\frac 1{\sqrt{N+1}}\sum_{n=0}^N|n\rangle _N\otimes |n\rangle
_N 
\end{equation}
And it also satisfies $(J_{1z}-J_{2z})|\Psi \rangle =0$. 

Now we consider the entanglement of two qubits, which belong to 
different ensembles. we first identify the 
two-qubit reduced density matrix : 
\begin{equation}
\rho _{12}=\left( 
\begin{array}{llll}
v & 0 & 0 & u^{*} \\ 
0 & w & 0 & 0 \\ 
0 & 0 & w & 0 \\ 
u & 0 & 0 & v
\end{array}
\right)  \label{eq:rhogood}
\end{equation}
in the basis $\{|00\rangle ,|01\rangle ,|10\rangle
,|11\rangle \}$,  which can be represented by
\begin{eqnarray}
v &=&\frac 14\left( 1\pm 2\langle \sigma _{1z}\rangle +\langle \sigma
_{1z}\sigma _{2z}\rangle \right) ,  \nonumber \\
w &=&\frac 14\left( 1-\langle \sigma _{1z}\sigma _{2z}\rangle \right)=
\frac 14-\frac{\langle J_{1z}J_{2z}\rangle }{N^2},
\label{eq:para} \nonumber\\
u &=&\frac 14(\langle \sigma _{1x}\sigma _{2x}\rangle -\langle \sigma
_{1y}\sigma _{2y}\rangle +i2\langle \sigma _{1x}\sigma _{2y}\rangle )\nonumber\\
&=&\frac{\langle J_{1+}J_{2+}\rangle }{N^2}.
\end{eqnarray}

The concurrence is given by
\begin{eqnarray}
{\cal C}&=&2\max \{0,|u|-w\} \nonumber\\
&=& 2\max \{0,\frac{\langle J_{1+}J_{2+}\rangle +\langle
J_{1z}J_{2z}\rangle }{N^2}-\frac 14\}  
\end{eqnarray}
The expectation values of $J_{1+}J_{2+}$ and $J_{1z}J_{2z}$ are readily obtained in the
state $|\Psi\rangle$, and we find that ${\cal C}=1/N$. The pair of particles is in an 
entangled state. If the ensembles are really macroscopic, as in \cite{Polzik}, the entanglement
is, however, very weak.
\section{Conclusions}

The purpose of this paper has been to point out that multi-particle
entanglement quite typically implies pairwise entanglement within the
sample. We showed that the two-particle density matrix is readily expressed
in terms of expectation values of collective operators, in the case of
symmetrical states of the many-particle system, and we provided the value of
the concurrence for a number of examples. These results confirmed and
generalized results obtained, e.g., on the pairwise entanglement in systems
with definite $(N=3,4)$ numbers of particles.

The entropy of formation, and the very issue of entanglement, is highly
non-trivial for situations dealing with more than two particles, and for
mixed states of systems with dimensions higher than 2. Studying and
optimizing the two-particle concurrence in systems with many particles may
be a useful way to learn about the more complicated case.

\acknowledgments

This work is supported by the Information Society Technologies Programme
IST-1999-11053, EQUIP, action line 6-2-1.

\end{document}